\documentclass[sn-mathphys]{sn-jnl}
\usepackage{graphicx}
\usepackage{hyperref}
\usepackage{url}
\usepackage{mhchem}

\jyear{2022}%

\begin{document}

\title[AutoMat]{AutoMat: Automated Materials Discovery for Electrochemical systems}


\author[1]{\fnm{Emil} \sur{Annevelink}}\email{eannevel@andrew.cmu.edu}
\equalcont{These authors contributed equally to this work.}
\author[1]{\fnm{Rachel} \sur{Kurchin}}\email{rkurchin@cmu.edu}
\equalcont{These authors contributed equally to this work.}
\author[2]{\fnm{Eric} \sur{Muckley}}\email{emuckley@citrine.io}
\author[1]{\fnm{Lance} \sur{Kavalsky}}\email{lkavalsk@andrew.cmu.edu}
\author[2]{\fnm{Vinay I.} \sur{Hegde}}\email{vhegde@citrine.io}
\author[1]{\fnm{Valentin} \sur{Sulzer}}\email{vsulzer@andrew.cmu.edu}
\author[1]{\fnm{Shang} \sur{Zhu}}\email{shangzhu@andrew.cmu.edu}
\author[1]{\fnm{Jiankun} \sur{Pu}}\email{jiankunp@andrew.cmu.edu}
\author[3]{\fnm{David} \sur{Farina}}\email{farina.d@northeastern.edu}
\author[4]{\fnm{Matthew} \sur{Johnson}}\email{mattsj@mit.edu}
\author[5]{\fnm{Dhairya} \sur{Gandhi}}\email{dhairya@juliacomputing.com}
\author[1]{\fnm{Adarsh} \sur{Dave}}\email{ardave@andrew.cmu.edu}
\author[1]{\fnm{Hongyi} \sur{Lin}}\email{hongyili@andrew.cmu.edu}
\author[4]{\fnm{Alan} \sur{Edelman}}\email{edelman@mit.edu}
\author[6]{\fnm{Bharath} \sur{Ramsundar}}\email{bharath@deepforestsci.com}
\author[2]{\fnm{James} \sur{Saal}}\email{jsaal@citrine.io}
\author[5]{\fnm{Christopher} \sur{Rackauckas}}\email{chris.rackauckas@juliacomputing.com}
\author[5]{\fnm{Viral} \sur{Shah}}\email{viral@juliacomputing.com}
\author[2]{\fnm{Bryce} \sur{Meredig}}\email{bryce@citrine.io}
\author*[1]{\fnm{Venkatasubramanian} \sur{Viswanathan}}\email{venkvis@cmu.edu}

\affil[1]{\orgname{Carnegie Mellon University}, \orgaddress{ \city{Pittsburgh}, \state{PA}, \country{USA}}}
\affil[2]{\orgname{Citrine Informatics}, \orgaddress{\city{Redwood City}, \state{CA}, \country{USA}}}
\affil[3]{\orgname{Northeastern University}, \orgaddress{\city{Boston}, \state{MA}, \country{USA}}}
\affil[4]{\orgname{Massachusetts Institute of Technology}, \orgaddress{\city{Cambridge}, \state{MA}, \country{USA}}}
\affil[5]{\orgname{Julia Computing}, \orgaddress{\city{Somerville}, \state{MA}, \country{USA}}}
\affil[6]{\orgname{Deep Forest Sciences}, \orgaddress{\city{Fremont}, \state{CA}, \country{USA}}}

\abstract{
Large-scale electrification is vital to addressing the climate crisis, but several scientific and technological challenges remain to fully electrify both the chemical industry and transportation. In both of these areas, new electrochemical materials will be critical, but their development currently relies heavily on human-time-intensive experimental trial and error and computationally expensive first-principles, meso-scale and continuum simulations. We present an automated workflow, AutoMat, that accelerates these computational steps by introducing both automated input generation and management of simulations across scales from first principles to continuum device modeling. Furthermore, we show how to seamlessly integrate multi-fidelity predictions such as machine learning surrogates or automated robotic experiments ``in-the-loop''. The automated framework is implemented with design space search techniques to dramatically accelerate the overall materials discovery pipeline by implicitly learning design features that optimize device performance across several metrics.  We discuss the benefits of AutoMat using examples in electrocatalysis and energy storage and highlight lessons learned.
}

\maketitle

\section{Introduction/Background}
\subsection{Electrochemistry and Climate Change}
Electrification of virtually every economic sector is critical in the fight against climate change, as it will enable society to rely on carbon-free energy sources such as solar and wind~\cite{Davis2018}. While these carbon-free electricity generation technologies are already relatively mature and scalable, increased performance and reduced cost of less-mature electrochemical technologies is currently the bottleneck to increased electrification. Two critical areas where electrochemical transformations can help are in energy storage and chemical synthesis. Many are already familiar with phenomena such as range anxiety (not to mention sticker shock) that hinder adoption of electric vehicles (EVs) \cite{krishnaUnderstandingIdentifyingBarriers2021}. Improvements in both energy and power capacity (as well as continued reduction in cost) of storage technologies such as lithium-ion batteries are necessary, not only to increase EV adoption, but to enable electrification of more energy-intensive modes of transportation such as trucking~\cite{e_truck} and aviation~\cite{e_flight}, as well as for demand smoothing on a grid dominated by variable renewable sources.

Perhaps less familiar than energy storage challenges are those posed to electrification of the large swaths of the chemical industry that rely on extreme conditions (heat and pressure) produced by burning fossil fuels for large-scale synthesis of chemicals that are essential for fertilizers~\cite{RS_ammonia}, steel~\cite{REN2021}, cement~\cite{andrewGlobalCO2Emissions2018} and other aspects of modern life that many of us take for granted. Electrochemistry represents a propitious opportunity to provide activation energies of reaction via voltage instead of via high temperatures and pressures. In principle, 1 V of electrical potential can ``replace'' $1\text{ eV}/k_{\text{B}}\simeq 11,500$ K of temperature. The engineering challenge then becomes how to efficiently direct this energy towards the reactions that we want, while suppressing competing unwanted and parasitic processes.

\subsection{Traditional Materials Design}
To realize the promise of electrochemistry to address storage and electrification challenges while also meeting demanding technoeconomic targets, novel materials are needed.

\begin{figure}
    \centering
    \includegraphics[width=\textwidth]{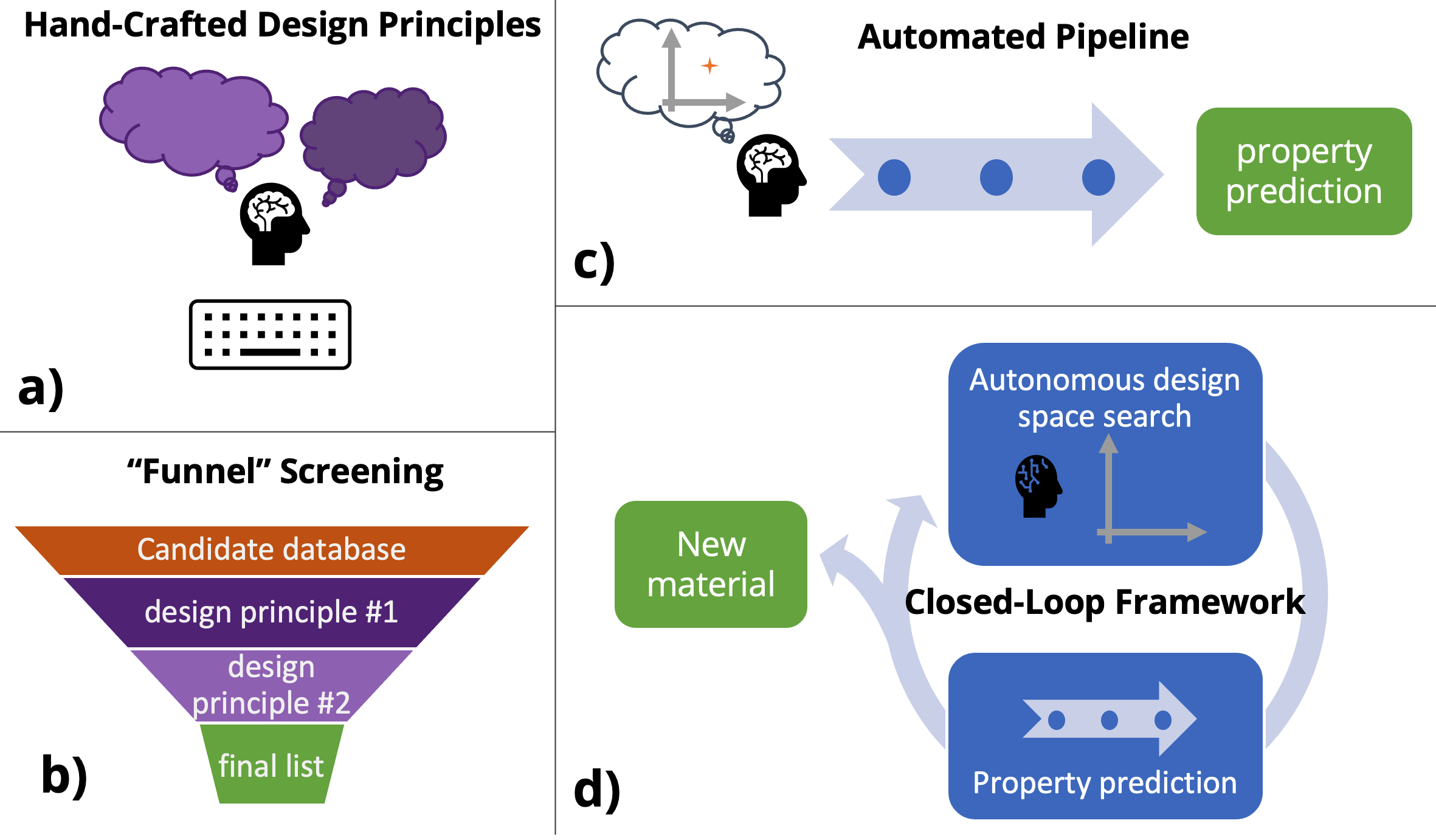}
    \caption{a) Traditional materials design, with researcher fully ``in the loop'' crafting strong design principles to guide ``manual'' computational screening. b) ``Funnel'' screening: a large initial list of candidates (usually from an open-source database) is progressively narrowed by more stringent search criteria, often analogized to a ``funnel'' approach. c) Automated pipeline: end-to-end performance prediction given an input material.  d) Closed-loop framework: System autonomously selects and evaluates candidates.}
    \label{fig:screening}
\end{figure}

Typically, materials discovery has relied on ``design principles'' that distill experience and intuition into guidelines that allow for designing and optimizing solutions (Figure~\ref{fig:screening}a).
We define a design principle (DP) as a heuristic that enables connection of a material property or descriptor to performance measures of interest such as efficiency, lifetime, cost etc.
The use of DPs allows researchers to quickly narrow down the design space to make it tractable for an experiment or simulation.
However, they normally fall short of being able to explicitly predict target device properties, instead serving to predict trends in performance.
These DPs can be derived from experimental measurements and/or simulations.
In particular, density functional theory (DFT) has become a popular tool for creating DPs for new materials \cite{hautier_computer_2012}, and has contributed to the proliferation of high-throughput systematic computational materials screening approaches (often analogized to a funnel, see Figure~\ref{fig:screening}b), wherein a large number of candidates is whittled down by a series of (typically increasing in computational cost) criteria. Such screening approaches have been enabled by vast databases and associated API's made available by efforts such as Materials Project~\cite{Jain2013}, OQMD~\cite{Saal2013}, and AFLOW~\cite{Taylor2014}. 

In catalysis, an overarching DP is the Sabatier principle, positing a sort of ``Goldilocks range'' of binding energies that lead to optimal reaction rates due to the balance between forward and reverse reaction rates\cite{medfordSabatierPrinciplePredictive2015}.
The Sabatier principle gives rise to an ``activity volcano'' and is widely used to design new catalytic materials by finding catalysts with binding energies that place them near the peak. However, directly computing or measuring these binding energies can be prohibitive, motivating a search for more accessible descriptors.

Experimental studies can probe catalytic activity to find correlations between measured properties and device performance.
For example, in Ir-based catalyst systems for the oxygen evolution reaction, both the surface concentrations of Ir$^{III+}$ and OH are correlated to the volcano peak, enabling screening of Ir-Ni catalysts based on ex-situ measurements \cite{spori_experimental_2019}.
However, with the increasing availability of HPC resources, descriptors from electronic structure calculations have become easier and faster to compute, leading to joint experimental-computational works to establish accurate descriptors \cite{hammer_dband_1995, jacobs_assessing_2019}.
DFT also enables direct computation of binding energies, at least in idealized systems. This has been done for a wide range of electrocatalytic reactions including hydrogen evolution reaction \cite{laursen_electrochemical_2012}, oxygen reduction and evolution reactions \cite{man_universality_2011}, and \ce{CO2} reduction reaction \cite{varela_electrochemical_2019}. 
However, such direct computation remains infeasible for many realistic situations, such as cases where interactions between different adsorbates substantially influence binding energies \cite{capdevila-cortada_performance_2016}.

In battery materials design, there is not a single overarching DP analogous to the Sabatier principle in catalysis.
Instead, the structure or composition of a single component -- such as the anode, solid-electrolyte interphase (SEI), electrolyte, or cathode -- is correlated to that component's performance, which is then correlated to the overall device performance.
One such correlation is between the formation of dendrites at the anode-electrolyte interface and reduced cycle life of lithium-ion batteries through the formation of internal shorts and electrolyte dry-out \cite{linRevivingLithiumMetal2017}.
Thus, descriptors that predict or suppress dendrite growth are crucial in designing long-lived batteries, particularly lithium-metal batteries.
One promising approach is developing artificial SEI's (ASEI) that facilitate creating more uniform stripping and plating at the anode \cite{thompson_stabilization_2011,thanner_artificial_2020}. 
Three experimental DPs have been formulated for developing new ASEI materials including (1) mechanical stability, (2) spatially uniform \ce{Li+} transport, and (3) chemical passivation \cite{yu_design_2020}. 
These were then used to qualitatively evaluate different electrolyte classes against one another, showing the effectiveness of polymeric multifunctional ASEIs \cite{yu_dynamic_2019}.

Unfortunately, formulating DPs is time-consuming and, although effective, is limited by the time researchers can devote to elucidating them.
To go beyond ``hand-crafted'' DPs, there need to be tools that can evaluate device performance directly from material structure.
Experimentally, this requires having an automated framework (Figure \ref{fig:screening}c) that performs all tests without human intervention.
Computationally, this requires multi-scale theory with seamless handoff of outputs from one scale to the inputs of the next, allowing for hierarchical simulation.
Both rely on the integration of research infrastructure to drastically accelerate the pace of hypothesis testing.

\subsection{Closed-loop Automated Frameworks}

With an automated, end-to-end property evaluation, simulated or experimental, the next step to go from a ``function evaluation'' framework to a materials design platform is to connect the two ends; namely, to let the result of one property evaluation inform where in the domain to look next, such that a design space search can happen autonomously (via active learning) in a closed-loop framework (Figure \ref{fig:screening}d). 
The design space search consists of two pieces: a global surrogate model and an acquisition function. 
The global surrogate models the full end-to-end pipeline, can be inexpensively retrained upon receipt of new data, and generates uncertainty estimates alongside its predictions. Common architecture choices are Gaussian processes~\cite{rasmussenGaussianProcessesMachine2004} and random forests~\cite{Breiman2001}. The acquisition function chooses the next point in the design space to evaluate, balancing ``exploration'' (visiting regions of the space that are underexplored/highly uncertain) with ``exploitation'' (evaluating likely optima of the performance function) to autonomously discover novel materials that maximize performance. 
Importantly, since the optimization contains all the evaluation information, it can leverage not just strong descriptors but also weak descriptors to form high-dimensional DPs not readily identifiable by humans.

Closed-loop frameworks incorporating automated experimentation have proliferated in recent years, encompassing optimization of the crystallinity~\cite{Li2020} or hole mobility~\cite{MacLeod2020} of thin films, efficiency~\cite{Du2021} or aging properties~\cite{Zhao2021} of emerging photovoltaic materials, mechanical properties such as hardness and wear resistance of metallic glasses~\cite{Sarker2022} or toughness of additively manufactured structures~\cite{Gongora2020}, batteries device performance~\cite{Attia2020}, battery materials~\cite{otto1,otto2,clio} and a variety of micro- and nanoscale synthesis processes~\cite{Nikolaev2016, Chang2020, Kusne2020, Fong2021, Tao2021, Wahl2021, Siemenn2022}. There are comparatively few cases incorporating closed-loop frameworks for purely in silico design tasks~\cite{Herbol2018, Flores2020, Bolle2020, Antono2020, Menke2021}, likely due to the widespread adoption of the ``funnel'' screening approach described above. 

Our framework, described in detail below, is among the first autonomous closed-loop materials design systems capable of incorporating both physics-based modeling and automated experimentation along with machine learning models as both component surrogates and as global surrogates for sequential-learning-guided acquisition functions for materials discovery.

\section{AutoMat}
\subsection{Overview}
 
\begin{figure}
    \centering
    \includegraphics[width=3.5in]{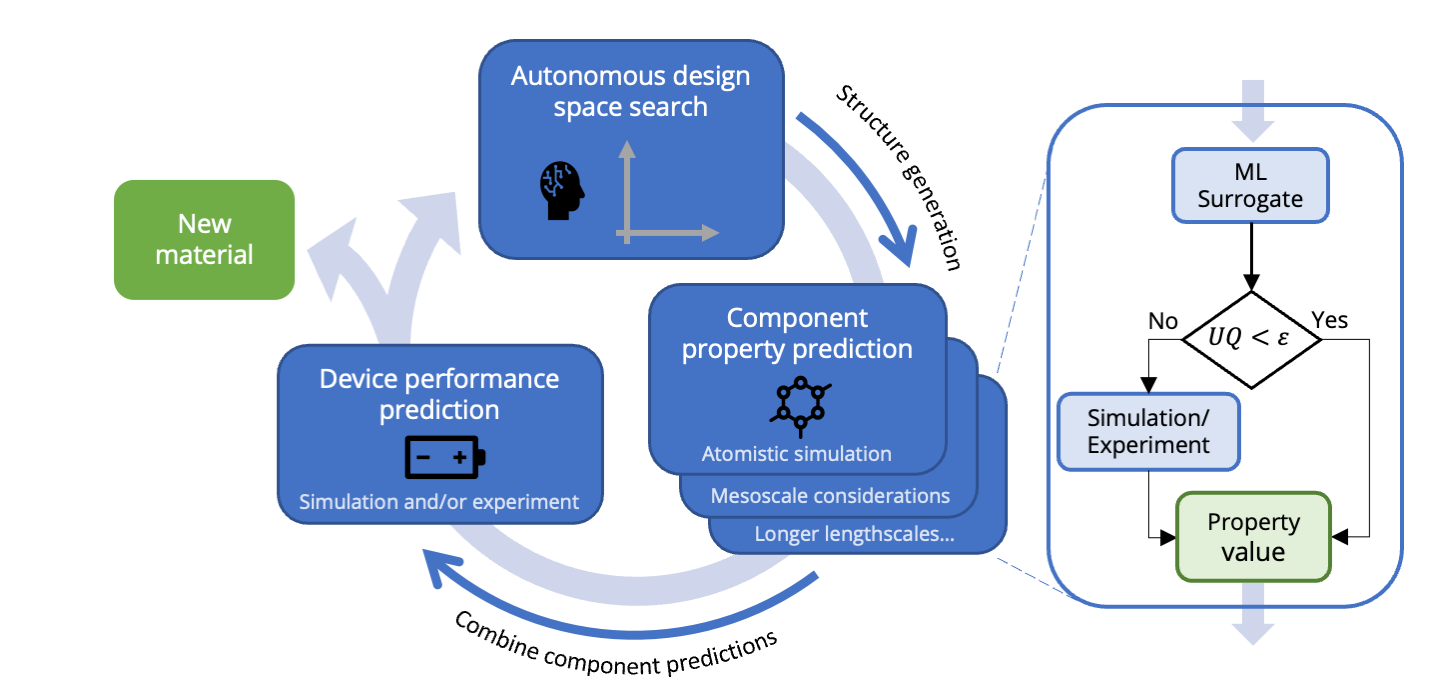}
    \caption{Our end-to-end structure device performance evaluator using multi-fidelity prediction (ML Surrogate and Physics Simulation) and experiment in the loop design philosophy to discover new materials.}
    \label{fig:designphilosophy}
\end{figure}

AutoMat is a modular framework for materials design comprising a reconfigurable end-to-end property evaluation system as part of a closed-loop system utilizing a global surrogate for design space search.
We note that many of the individual components are not new (though we have made modifications/upgrades in some cases), but by combining pieces together in an automated fashion, including multiple different levels of theory within a property calculation, we can dramatically accelerate performance prediction and enable faster design space searching.
In essence, we are ``not reinventing the wheel, but building a highway."

Figure \ref{fig:designphilosophy} shows a high-level schematic of the framework, and the next section describes applications.
The active learning loop requires a structure generator, a set of property/performance evaluators, and autonomous design space search.
The structure generator is used to translate a design choice to an input for the property evaluators.
Property evaluation can be achieved by a physical simulation, an ML surrogate of that simulation, and/or experimental measurement.
Each operate at different fidelities, which typically exhibit an accuracy-cost trade-off (financial, computational, and/or temporal).
For multi-scale pipelines, each lengthscale has its own property evaluator, which are typically chained together from smallest (atomic) to largest (device) scales, with each longer scale requiring input from shorter ones. 
The output of the final step is a device performance prediction, which is returned to the global surrogate for retraining, and to the acquisition function to choose a subsequent point in the design space for evaluation.
If the device performance prediction exceeds a set target, then we have discovered a new material for further investigation!

Given that acceleration of materials design is a primary goal of designing such a framework, quantifying this acceleration is key. There are two broad categories that contribute to the bulk of the acceleration. The first is acceleration due to automation of property evaluation and elimination of human-associated lag times. These speedups will be detailed in subsequent sections. The second is the reduction in total number of full candidate evaluations required arising from efficient design space search.
The number of candidate evaluations is decreased by using a global surrogate with uncertainty estimates coupled with an acquisition function within the FUELS framework~\cite{ling2017high}.
For both applications below, we use the ``maximum expected improvement'' \cite{ling2017high} acquisition function to accelerate electrocatalyst search by 3$\times$ and liquid electrolyte search by 15$\times$.

The following sections showcase two implementations of the AutoMat framework.
For each example, we lay out the vision and describe the current status, showing how the framework is modular and allows for a bare-bones implementation that can be subsequently extended. In both cases, the currently-implemented workflow is of a fixed ``topology,'' and planned improvements involve introducing ``branches'' to the workflow with automatic decision-making between, for example, relying on the results of a component surrogate, or choosing to run a higher-fidelity evaluator if the predicted uncertainty of the surrogate is above some threshold.

\subsection{Application to nitrogen reduction electrocatalysis}
Ammonia production via the Haber-Bosch process (the vast majority for use in fertilizer) is the largest source of carbon dioxide emissions of all chemical production, exceeding the next-largest emitter (ethylene) by more than a factor of two and corresponding to approximately 1.8\% of global \ce{CO2} emissions~\cite{RS_ammonia}. Thus, decarbonizing ammonia production by finding a viable electrocatalyst for the nitrogen reduction reaction (NRR) represents an enormous potential climate impact from materials development. However, achieving both high activity and sufficient selectivity in electrochemical NRR has proven challenging~\cite{Suryanto2019}.

One emerging materials class that has shown preliminary promise for a variety of heterogeneous catalysis problems is single-atom alloys (SAA's)~\cite{Giannakakis2018}. SAA's can be thought of as extremely dilute alloys where the alloying element disperses on the catalyst surface to form single-atom active sites. Their unusual electronic structure may allow them to circumvent the commonly-observed linear scaling relations between various reaction intermediates and access higher catalytic activities~\cite{Hannagan2020-uk}. The large combinatorial space of potential SAA materials is ideal for exploration by a closed-loop framework such as AutoMat.

In this application, there are two lengthscales at play: first, the atomistic scale, where we need to understand structures and energetics of adsorption of all relevant steps in the NRR mechanism. Once these are computed for a given candidate, we can solve a set of microkinetic modeling equations to predict performance metrics governing activity of the catalyst and its selectivity for NRR compared to competing pathways. The atomistic lengthscale is by far the most computationally expensive step (generally requiring $\mathcal{O}(10)$ DFT calculations per candidate), making it ideal for surrogatization so that full calculations are only run when needed (i.e. the component surrogate is uncertain).

Currently, we have implemented an automated pipeline for DFT calculations of nitrogen adsorption energies.
Once adsorption energies are obtained, catalyst activity is predicted using a previously-published activity volcano~\cite{nrrscaling}. 
Work on extending this bare-bones implementation through additional property evaluators is detailed in the SI.

\begin{figure}
    \centering
    \includegraphics[width=\linewidth]{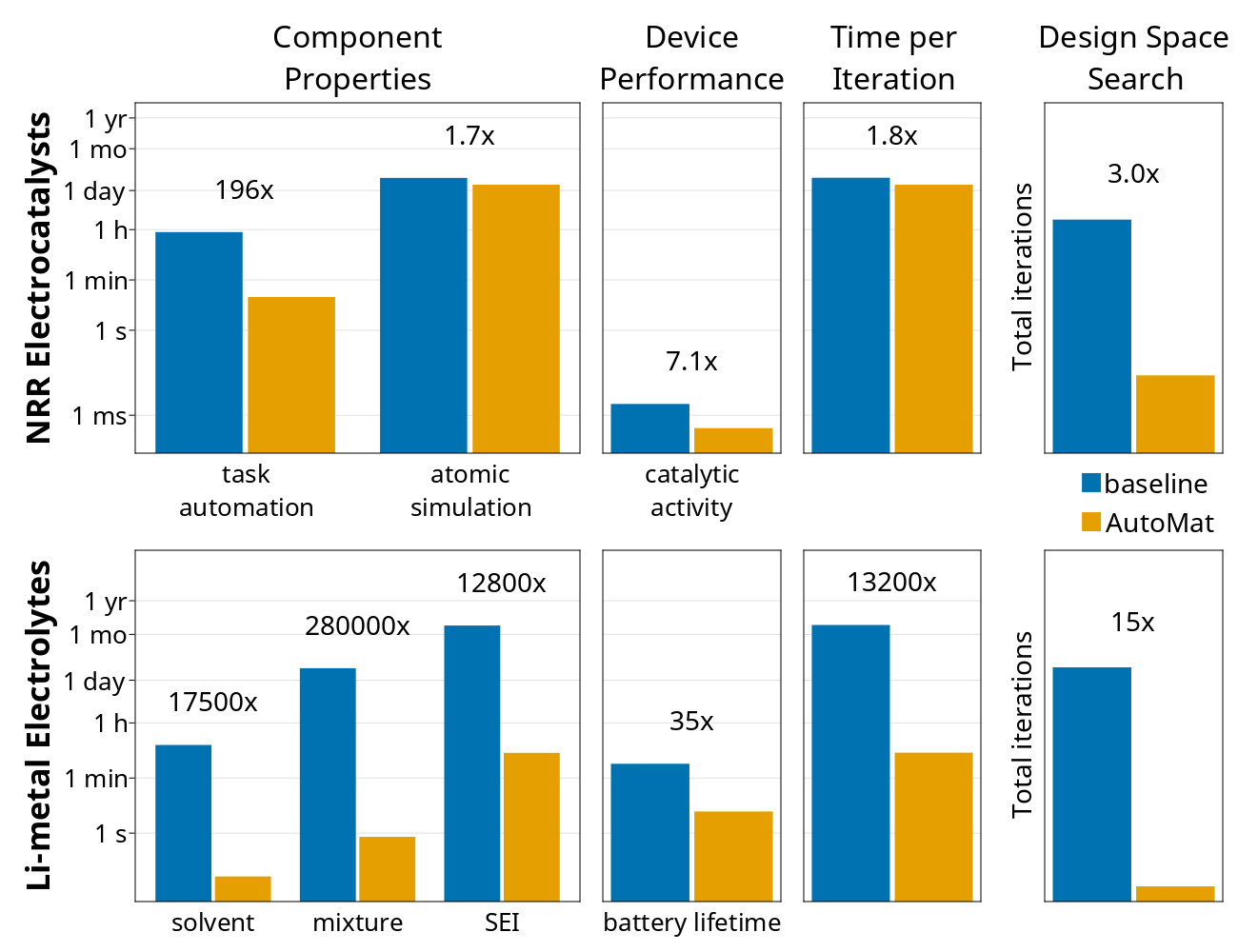}
    \caption{Acceleration of individual component models developed for catalysis (above) and batteries (below).}
    \label{fig:acceleration}
\end{figure}

Timing estimates of the individual workflow components relative to their traditional counterparts are reported in Figure \ref{fig:acceleration} and quantify the benefits of the AutoMat framework. 
The first advantage stems from the automation of the tedious and repeated tasks needed for a computational study such as structure generation, job management, and data compilation.
Another advantage is improving calculation runtime.
We estimate that better initial adsorbate placement and calculator settings can reduce the time per adsorbate energy calculation by 1.7$\times$.

\subsection{Application to liquid electrolytes for lithium metal batteries}

Utilizing a lithium-metal anode in lithium-ion batteries promises a doubling of energy density \cite{liu_advancing_2018}. 
Lithium reactivity, however, is a challenge for the standard electrolyte species used in conventional graphite anode cells, leading to issues such as dendritic internal shorts (discussed above).
Designing a new electrolyte mixture could resolve these shortcomings through a stable SEI that prevents electrolyte decomposition and uniformly plates and strips lithium \cite{yu_molecular_2020}.
Although promising strides have been made over the past few years \cite{hobold_moving_2021}, higher performance is still demanded in many applications, necessitating continued exploration of the large design space of electrolyte molecules and blends.

To design a new electrolyte, we use AutoMat to chain property calculators into a multi-scale simulation pipeline that starts with solvent and salt molecules, calculates their electronic and mixture properties, estimates their reactivity with a lithium surface, and finally predicts battery performance.
At the microscopic scale, the electronic properties of potential solvent and salt molecules inform their stability and solvation using high-throughput quantum chemistry (QC) methods.
To alleviate the cost of expensive QC calculations, a component surrogate will be trained on previous QC data.
One lengthscale up, mixture properties can be evaluated at three fidelities -- molecular dynamics simulations, experiment, and ML component surrogate -- where the AutoMat framework will automatically determine when to use each based on uncertainty criteria.
Next, the reaction network of electrolytes with lithium surfaces is generated to determine the decomposition of salts and solvents with the lithium surface to predict electrolyte decomposition and interface resistances.
Finally, solvent and salt properties, mixture properties, and electrolyte reactivity will be passed to battery simulations to directly associate the electrolyte composition with device performance.

The current implementation of AutoMat for batteries is a bare-bones connection of mixture properties and device performance.
We use our robotic test-stand Clio~\cite{clio} to mix electrolyte blends and determine their density, conductivity, and viscosity.
A component surrogate predicts electrolyte properties to accelerate property evaluation.
The electrolyte conductivity is input into the battery modeling software PyBaMM~\cite{pybamm} to evaluate the cycle life of each electrolyte.
The cycle life is optimized to propose a new electrolyte formulation.
Additional tools, such as those reported in the SI, can readily be incorporated according to the vision to increase the accuracy and generalizability of electrolyte predictions.

The acceleration of different workflow components is compared against baselines and shown in Figure \ref{fig:acceleration}.
The accelerations cover both components we have implemented in the workflow and those in development as described in the SI.
The solvent property baseline is a Psi4~\cite{psi4} QC calculation, which we accelerate with a DeepChem~\cite{deepchem} surrogate. 
The mixture properties are calculated using the OPLS-AA\cite{oplsaa} force field in LAMMPS~\cite{lammps}, which is surrogated with a permutationally invariant neural network on the individual components.
The surrogates for both solvent and mixture properties virtually eliminate the computational cost of property evaluation (we note that there is a ``fixed cost" to generate training data). Work is ongoing to obtain well-calibrated uncertainty estimates to assess when the full simulation should be run.
Since we upgraded Reaction Mechanism Generator (RMG)~\cite{rmg2} to fill a simulation gap, there isn't a direct baseline other than human evaluation and intuition, which we estimate as taking 20 minutes per species considered and 1 hour per reaction considered for each of the 900+ species and 1300+ reactions examined in a test RMG run.
The battery lifetime models are accelerated
through solutions to the degradation modeling~\cite{sulzer2021accelerated}.
Combined, these give a predicted acceleration of approximately thirteen thousand.

\section{Perspective and Outlook}
A variety of interesting challenges arise when endeavoring to construct a fully automated closed-loop system like AutoMat. One class of such problems emerges at the interface between the various length and time scales. Most computational researchers receive intensive training in their specialty technique, be it first-principles simulations, continuum modeling, or data-driven approaches.
However, developing a multiscale tool requires the ability to bridge scale gaps. This ability emerges from interactions between practitioners at various scales. The group must understand precisely what inputs are needed for a given computation and determine how to obtain them from the previous one. And, of course, ``driver'' code needs to be written to parse those outputs, format them, and pass them on.
One example of this arose in connecting the molecular dynamics to PyBaMM models. 
When studying liquid electrolytes, we hadn't initially considered calculating the activity coefficients using molecular dynamics, but the results of the PyBaMM models depend sensitively upon these parameters, so computing them accurately is crucial. 

Facilitating the seamless communication that leads to identification and resolution of issues like this also represents a logistical challenge in an international collaboration. When the project first began, we had extensive discussions to build consensus on platforms for collaboration (primarily GitHub, Slack, and Zoom) as well as team and subteam structure and meeting schedule. With teammates across continents, time zones are a challenge and so we have a single one-hour meeting for the whole team each week (with an ``off week'' once per month), and the majority of synchronous communication happening in smaller meetings. We found that spending the time to solicit input and come to agreement on these structures facilitated buy-in and was crucial to building strong working relationships.

AutoMat is a modular framework that enables modular, extensible, multi-fidelity implementations to easily balance evaluation cost and speed.
It extends traditional design-principle-driven materials discovery by utilizing both strong and weak descriptors to accelerate design space search.
Through case studies in catalysis and energy storage, we show component accelerations of orders of magnitude and design space search acceleration of 3-15x.  Furthermore, AutoMat provides a method for seamless integration of experiment and computation into a high-throughput workflow.  We believe that AutoMat is poised to play an important role in the electrification transformation.

\subsubsection*{Acknowledgments}
The information, data, or work presented herein was funded in part by the Advanced Research Projects Agency-Energy (ARPA-E), U.S. Department of Energy, under Award Number DE-AR0001211. The views and opinions of authors expressed herein do not necessarily state or reflect those of the United States Government or any agency thereof.

\subsubsection*{Conflict of Interest}
On behalf of all authors, the corresponding author states that there is no conflict of interest.

\bibliography{refs}

\appendix

\section{Planned Improvements to initial AutoMat implementations}

\subsection{Nitrogen Reduction Reaction}

The bare-bones implementation performs a full DFT calculation to find the adsorbate energy and predicted device performance from a previously published activity volcano.
Currently, every performance prediction requires a full set of DFT calculations for the nitrogen adsorption energies, which is computationally demanding. We are working on developing a lower-fidelity machine-learned surrogate to compute binding energies, including uncertainty quantification in order to assess whether a full DFT calculation needs to be done. We estimate, accounting for the need to generate training data, that the introduction of these component surrogates will provide up to another 4-5x acceleration.

Currently, in generating the DFT structures, the adsorbates are placed based on pre-defined heuristics with corresponding calculation defaults. Looking ahead, chemically informed placement and choice of calculator settings are estimated to have the potential for an additional 1.1-2.3$\times$ speedup. 

We also have the capability to predict turnover frequency (TOF) via solving a full microkinetic model implemented in ReactionMechanismSimulator.jl~\cite{RMS}, which we plan to integrate into the automated pipeline shortly. This improves over the current approach of using a thermodynamic activity volcano in two ways: 1) TOF is an experimentally accessible quantity~\cite{Norskov_2005, Hansen2014, Liu2017-sg}. 2) We avoid the use of linear scaling relations which may not be as accurate for SAAs~\cite{Hannagan2020-uk}.

\subsection{Liquid Electrolytes for Lithium Ion Batteries}

\subsubsection{Planned Improvements}
The bare-bones implementation can optimize electrolyte composition to maximize battery cycle life.
However, there are many components that can increase the accuracy of the limited implementation in accordance to the vision described in the main text.
At the microscopic scale, we have developed a DeepChem~\cite{deepchem} machine learning surrogate trained on open-source databases, which we supplement with an automated quantum chemistry (QC) pipeline implemented using Psi4~\cite{Smith2020} through the QCEngine~\cite{Smith2021} interface.
Uncertainty estimates of the surrogates are generated using an ensembling approach to automatically decide when a new QC calculation was required.
Currently, this component can be run from our automated framework, but we are still working on how to attach the solvent properties to larger lengthscale simulations.

Mixture properties derived from molecular dynamics typically have suffered from limited accuracy.
We have recently developed a molecular dynamics workflow based on chemical physics to better match experimental techniques, which we are now working to automate the technique and ensure its reliability in other systems.
It can then be used in tandem with Clio~\cite{clio} and the mixture surrogates to determine electrolyte properties in-silico without the need to synthesize molecules directly.

The reactivity of the electrolytes is determined using the Reaction Mechanism Generator (RMG) \cite{rmg1, rmg2} to generate the decomposition pathway of solvent molecules in a lithium environment.
RMG was first designed to study the decomposition of hydrocarbons.
To study electrolyte decomposition required extending its reaction template library to include lithium.
The final piece is estimating the reaction kinetics to simulate the reaction rates and determine decomposition products.
Similar to the molecular dynamics pipeline, extending RMG to lithium chemistry was a major scientific endeavor, which we believe will greatly benefit the battery community.
We are now working on integrating RMG into the pipeline which requires automation and interpretation of its outputs into specific PyBAMM inputs.

\end{document}